\documentclass{aa}
\usepackage{epsfig,graphics,psfig,rotate}

\begin{document}

   \thesaurus{08.14.2; 09.10.1}
\def\xr#1{\parindent=0.0cm\hangindent=1cm\hangafter=1\indent#1\par}
\def\nv{GQ Mus }
\def\oneskip{\vskip\baselineskip}
\def\xr#1{\parindent=0.0cm\hangindent=1cm\hangafter=1\indent#1\par}
\def\am{$\arcmin$}
\def\as{$\arcsec$}
\def\cdt{$c\Delta t$}
\def\exo{{\sl Exosat }}
\def\ea{\it et al. \rm}
\def\ein{{\sl Einstein }}
\def\la{\raise.5ex\hbox{$<$}\kern-.8em\lower 1mm\hbox{$\sim$}}
\def\ha{H$\alpha$}
\def\ma{\raise.5ex\hbox{$>$}\kern-.8em\lower 1mm\hbox{$\sim$}}
\def\msol{M$_{\odot}$ }
\def\te{$T_e$}
\def\th{$\theta$}
\def\thd{$\theta D$}

\title{The emission nebula associated with V1974 Cyg:
a unique object?}
\author { R. Casalegno \inst{1,2},
 M. Orio \inst{1,3},
 J. Mathis \inst{3},
 C. Conselice \inst{3},
 J. Gallagher \inst{3},
 S. Balman \inst{4},
 M. Della Valle \inst{5},
 N. Homeier \inst{3},
 H. \"Ogelman \inst{6}\\}

\offprints{R. Casalegno}

\institute{Osservatorio Astronomico di Torino - Strada Osservatorio
 20, I\-10025 Pino Torinese (TO), ITALY \and email:
 casalegno@to.astro.it \and Department of Astronomy, University of
 Wisconsin, 475 N. Charter St., Madison WI 53706, USA
\and Department of Physics, Middle East Technical University,
In\"on\"u Bulvari, Ankara 06531, Turkey
\and
 Osservatorio Astronomico di Arcetri, Largo E. Fermi 5, Firenze, Italy
 \and Physics Department, University of Wisconsin, 1150 University
 Ave., Madison WI 53706, USA }

\titlerunning{Nebula associated with V1974 Cyg}
\maketitle

\abstract{ Through a program of narrow band imaging, we have observed
the changing structure of the H$\alpha$ emission line around Nova Cyg
1992 (V1974 Cyg) at regular intervals from 1996 to 1999. Between 1994
and 1996, the nebular boundary advanced to the southwest at nearly the
speed of light, implying that the nebula was created by an expanding
wave of radiation originating in the explosion interacting with
surrounding material. The expansion speed dropped to 0.35$c$ during
1996-1999. We have taken spectra of the nebula in 1998 and 1999.
Only Balmer lines are detected, no He I, [O~III], [O~II], [N~II],
or [S~II]. There is also no trace of the high excitation nova lines
(He II, NeV, etc).
The Balmer lines are unresolved in velocity (FWHM$\leq$100 km s$^{-1}$). 
These spectra show that the nebula is {\it not} a
reflection nebula, a conventional H~II region, or a shock involving
motions of the gas. The integrated H$\alpha$ luminosity 
of the nebula between 1996 and 1999 is in the range
$\simeq$1.3-2.2 $\times$ 10$^{35}$ erg s$^{-1}$.

The Balmer decrement is normal for recombinations of
a lightly reddened plasma.  The lack of forbidden
emission lines can only be understood if the electron temperature is
low. This condition results if the energies of the ejected
photoelectrons are shared among electrons, protons, and neutrals in a
partially ionized medium. The He~I lines are suppressed if the flash
ionizing spectrum is truncated at or below the He$^0$ ionization edge.
The ionized material is on the front face of
neutral sheets. The density is poorly determined, but is probably very
large ($\sim10^4$ cm$^{-3}$) in order to explain the
brightest region of the nebula. The dynamical timescale is about a
year and the recombination timescale of the same order. Bright patches
are observed to fade in these times. The energy required to ionize the
nebula is the bolometric luminosity of the nova for 30 days, smaller
than the time during which the temperature of the nova photosphere was
in the right range to produce the ionizing photons.

We have also undertaken sensitive surveys of H$\alpha$ nebulae around
recent novae but find no evidence of other such nebulae, so this type
of object must be rare.}

\keywords{Stars: novae, cataclysmic variables -- ISM: jets and outflows}
\section{Introduction}

Nova Cyg 1992 (V1974 Cyg) erupted in February 1992 and was the
brightest nova in the 24 years between V1500 Cyg 1975 and Nova Velorum
1999. It has been the object of detailed and frequent observations in
all wavelengths from radio to gamma rays. It reached visual maximum
V=4.4 on February 22.5 (Rafanelli et al. 1995).  The lightcurve was
characterized by time of decay by 2 magnitudes t$_2$=25 d (Kolotilov
et al. 1994) which implies, according to the maximum magnitude {\sl
vs.}  rate of decline relationship (Della Valle and Livio 1995), an
absolute magnitude at maximum of $V = -7.7$. Color excesses in the
range from $E(B-V)= 0.17$ (Rosino et al. 1996) to 0.35 (Paresce et al.
1995) give distances in the interval $1.6 - 2$ kpc (see also Chochol
et al. 1993). We adopt hereafter
d = 1.8 kpc. Immediately following the outburst, the photosphere was
at an effective temperature $\simeq$20000 K (see Hauschildt et
al. 1994). Then the photosphere shrunk at constant bolometric
luminosity while the ejecta were gradually becoming ionized.  The
effective temperature reached $\simeq$300000 K 252 days after the
outburst and a maximum T$\simeq$590000 K at day 511 (Balman et
al. 1998).  At this point the effective
temperature rapidly decreased to T $\simeq$ 380000 K at day 612, and
it was again as low as T $\simeq$ 20000 K after two more years (Shore
et al. 1996).  Not every nova appears as a supersoft X-ray source
after the outburst (the reason is not quite clear). N Cyg belongs to
the few novae in which the very hot central star emerged at
some point (see Orio et al. 2000).

A nova outburst, specially one followed by a supersoft X-ray phase, can
cause an expanding reflection nebula, or its ionizing radiation can
ionize the surrounding interstellar medium (ISM), including cold,
neutral remnants of former outbursts. This paper deals with the first
spectral and temporal observations of a transient extended
nebula surrounding a post-nova and its ejected shell.
We still do not know how common transient nebulae around classical
novae might be. The technical capabilities of modern instruments and
telescopes (fields of view of at least several arcminutes, with good
sensitivity) prompted us to start a search for nebulae around known
nova remnants. We were interested both in {\it reflection}
and {\it ionization} nebulae.

Only one transient nebula is surely known to have been
associated with a nova:
the first superluminal astrophysical object, GK Per. This
was a {\it reflection} nebula (another reflection
nebula is suspected to have been around N Sgr 1936 for
some time, Swope 1940). The spectacular
GK Per nebula was discovered
at Yerkes Observatory in Wisconsin in 1901 (Ritchey 1901),
six months after the nova outburst. Partial rings of reflected light
centered around the nova appeared at a maximum angular distance of
about 6\am{} from the nova, corresponding to approximately 3 pc.
These features were seen moving outwards from the nova at the apparent
speed of 11$\arcmin$ per year. Later, when the distance to the nova
was known, it was understood to imply an amazing expansion velocity of
4.7$c$. The fascinating story of the light echoes of GK Per, so much
resembling those of SN 1987a, and how difficult they were to
understand, can be found in popular science article by Felten
(1991) summarizing early observations and 
discussions by Ritchey (1901 and 1902) Perrine (1901a, 1901b, 1903)
and others, 
and the controversial issues they raised. The case of
GK Per remained unique among explosive variables until it was
reproduced in the light echo of SN 1987a.

For novae appearing as hot supersoft X-ray sources
in X-ray telescopes,  one might expect
an extended {\it ionization} nebula. The discovery of 
a ionization nebula would let us understand better the history of the system and
the presence and composition of remnant material from previous
outbursts.  One such
nebula, around CAL\,83, has been detected (Pakull \& Motch 1989;
Remillard et al. 1995), and the 
physics is described by Rappaport et al. (1994). However, these models
are not applicable to classical  novae because of the timescales involved.
Novae seem to turn into {\it transient} supersoft X-ray sources
lasting only for few years, rather than into steady sources like CAL 83
(see Orio et al. 2000 and references therein).

The chances to find nebulae around novae in the past were slim due to
the low sensitivities of the instruments and of the photographic
plates. Later, the odds reduced again because the use of small format
CCDs was preferred to photographic plates that have much larger fields
of view.  The last systematic search for novae nebulae we are aware of
was done by van den Bergh (1977) with negative results.  Therefore we
still do not know how common transient nebulae around classical novae
might be. V1974 Cyg, as a known supersoft X-ray source, was the ideal
candidate to start our search.

\section{The discovery of the nebula through imaging }

The nebula was discovered almost simultaneously by two groups in which
the coauthors of this paper were taking part, both through
spectroscopy and through imaging.  An extended nebulosity was
discovered scanning spectra taken in 1994-1995 perpendicular to the
direction of dispersion in the H$\alpha$ line (Rosino et al.
1996). The diameter of the upper boundary of the emission region was
about 3.0$\pm$0.2\am{} in July 1994, 3.5$\pm$0.1\am{} in May and
3.95$\pm$0.1\am{} in October 1995. Since this angular velocity of
expansion is consistent with the speed of light at the
estimated nova distance, Rosino et al. (1996) suggested that the
nebula was due to ISM excited by the ultraviolet radiation emitted
shortly after the February 1992 outburst. The nebula after the initial
heating was roughly symmetric.

\begin{figure}
\resizebox{\hsize}{!}{\includegraphics{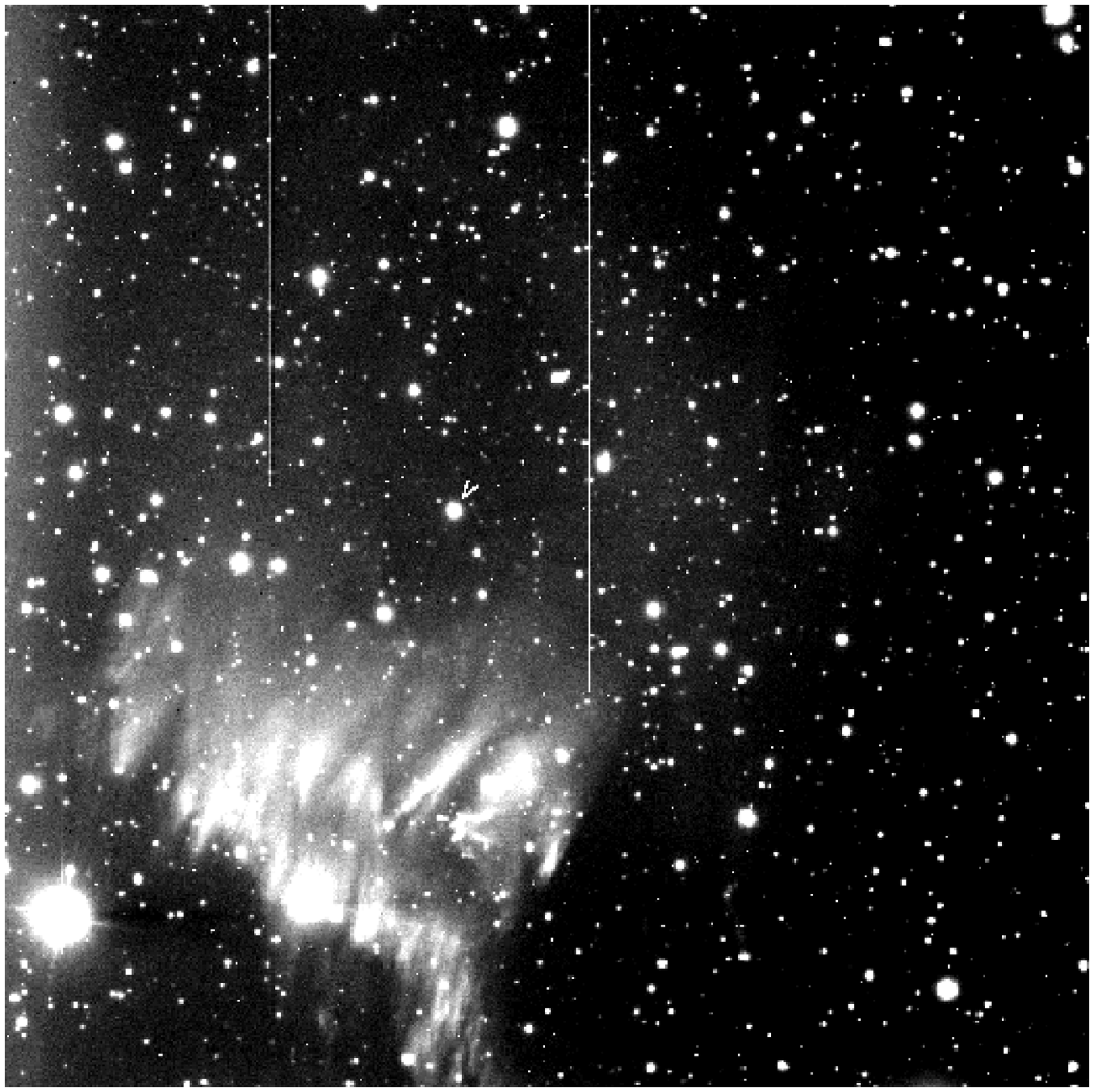}}
\resizebox{\hsize}{!}{\includegraphics{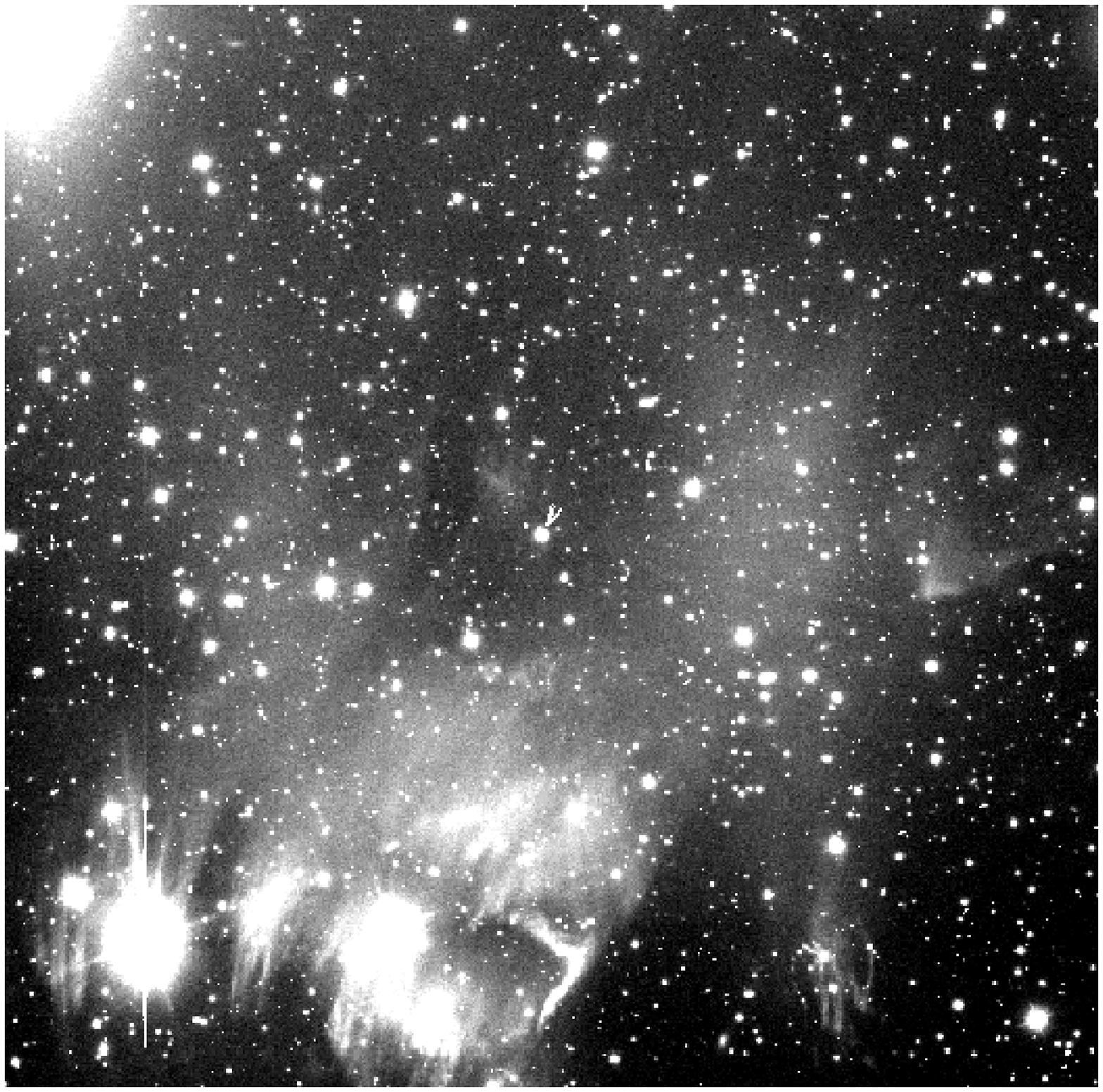}}
\caption{Top: the image of the nebula taken in June 1996.
Bottom: the Nebula in July 98. In both images  the nova
is indicated by the white arrow near the center of the images. The scale
of the images is 6.7 arcmin; North is up and East to the right.} 
\label{fig1}
\end{figure}

In 1996 the nebula was imaged 
using the WIYN telescope\footnote{The WIYN
Observatory is joint facility of the University of Wisconsin-Madison,
Indiana University, Yale University, and the National Optical
Astronomy Observatories.} at Kitt Peak equipped with 6.8\am{} field
CCD camera. The WIYN is a 3.5m f/6.5 telescope with an effective
diameter of ~3.2m (because of obstruction by the secondary mirror).
On 27 May 1996 a diffuse emission was detected in narrow band
H$\alpha$ around V1974. This emission appeared to extend 125$\arcsec$
north, 190\as{} south, 140\as{} west and 142\as{} east (Balman et
al. 1996; see also Fig.~\ref{fig1}). The nebula did not appear to be
filled but to have an elliptical cavity extending 100\as{} north,
21\as{} south, 35\as{} west, and 38\as{} east. The H$\alpha$
luminosity of the nebula had a peak in the southwest with what Balman
et al. (1997) described as long, thin, chevron like filamentary
structures (like in the latter image in Fig.~\ref{fig1}). We also 
imaged the nebula with an [O~III] ($\lambda$ 5008) narrowband filter
and did not detect any nebulosity, with an upper limit to the flux
$\sim$10$^{-13}$ erg cm$^{-2}$ s$^{-1}
,\:\sim10^{-3}$H$\alpha$. 

 On 16 June 1996, the H$\alpha$ nebula showed no substantial changes;
the narrowband \ha\ luminosity of the nebula in this second
observation was 2-3 times that of the nova, which also showed a shell
diameter of 0\as.8 with an extension to the northwest.  Very precise
determinations of the flux for the images of May and June 1996 were
not possible because the CCD was found to be non-linear at low
levels. However, in Table 1 we give an estimate for May 1997 which
should be correct to within 15\%. The method of measuring the flux is
discussed in detail in the next section. At a distance $\simeq$ 1.8
kpc (see Introduction), the size of the nebula implied that the bright
region had extended at a significant fraction of the speed of light at
the outburst. The rapid extension indicated a nebula soon after ``turn
on'' of the ionizing source, differing greatly from the nebula
discovered around CAL 83 (Remillard et al. 1995).

 Since 1995, the nebula seemed to have extended more towards South by
 May-June 1996 and to have lost its almost spherical symmetry.  The
 expansion in direction South-West continued in the following years of
 our monitoring (see Fig.~\ref{fig2}). The expansion southwards must
 have been 37\as{} between October 1995 and May 1996, approximately
 consistent with the rate of expansion in 1994-1995. 

The nova was monitored with the WIYN telescope from 1996 through the
end of 1999. Additional and complementary observations were performed
in 1997-1998 with the 1m REOSC reflector at the Turin Observatory in
Italy, using an 11\am{} field CCD camera.  Meanwhile, Garnavich \&
Raymond (1997) found that between October 1996 and May 1997 
the brightest emission
region, about 2\am{} southwest of the nova, appeared displaced by
10\as{} to the southwest while another feature appeared
 to have shifted in southeast direction by 30\as{} over a 200-day
 interval, consistently with a speed 0.5$c$ assuming a distance 
1.8 kpc. 

\begin{table}
\begin{tabular}{@{\hspace{.7cm}}c@{\hspace{.7cm}}|@{\hspace{.7cm}}
c@{\hspace{.7cm}}}
{\bf DATE} & {\bf Flux (erg cm$^{-2}$ s$^{-1}$)}\\
& \\
\hline
& \\
May 1997 & 4.56$\times 10^{-10}$ \\
July 1998 & 3.83$\times 10^{-10}$\\
July 1999 & 3.37$\times 10^{-10}$\\
November 1999 &5.7$\times 10^{-10}$\\
& \\
\hline
\end{tabular}
\caption{Fluxes between 1997 and 1999}
\label{tab1}
\end{table}

\section{Flux estimates and astrometry}

We observed the nova shell in H$\alpha$ (and in the R filter for
comparison) with the WIYN telescope and a {\it linear} CCD camera on
16 November 1996, 1 June 1997, 8 August 1997, 15 July 1998, 18 and 19
August 1998, 18 July 1999, 8 and 9 November 1999.  Fig.~\ref{fig1}
shows the H$\alpha$ nebula detected in July 1998. The field of view of
these images is 6.8 arcmin. On 15 September 1997 and 23 December 1997
we obtained additional large scale images (11 arcmin field of view)
with a CCD camera and the 1m REOSC reflector at the the Torino
Observatory. The REOSC images did not provide a S/N ratio good enough
to estimate the flux, but their large size allowed us to choose the
boundaries of the emission features for the flux integration. At the
WIYN telescope we imaged the nova field also with the [S~II] filter
(centered around $\lambda$6718 \AA) in August 1997, and did not detect
any emission. The upper limit is about 0.001 times the measured
H$\alpha$ intensity.


\begin{figure*}
\resizebox{12cm}{!}{\includegraphics{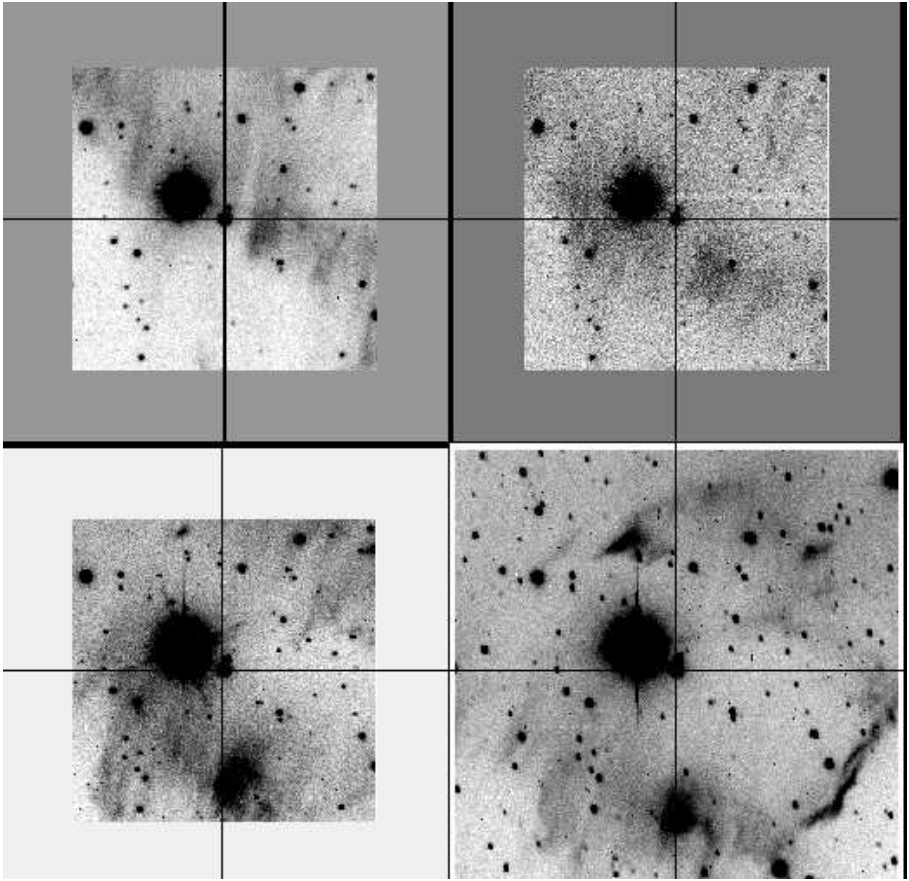}}
\hfill
\parbox[b]{55mm}{
\caption{The movement of the bright knot (on which the KPNO 4-m
spectra was taken) and the southern edge of the nebula as seen with
respect to a background star in June 1996 (upper left), June 1997
(upper right), August 1998 (lower left) and July 1999 (lower
right). The frame on the lower right is almost 1.5 arcmin
across. North is at the top, South at the bottom, East is right and
West is left.}
\label{fig2}}
\end{figure*}

We calibrated and analyzed the images using the IRAF software.  The
main problem was separating the contributions of the nebulosity from
the background and stellar light. To remove the latter, a PSF fitting
and subtraction was applied, with excellent results on non-saturated
stars.  The flux of saturated or non-subtracted stars was then
subtracted, setting an upper threshold for the average intensity of
the nebula pixels. The background was measured in sky zones of the
images that are far from the nebulosity. To determine the borders of
the nebula, we measured the gradient of the intensity in 10 x 10 pixel
windows, and defined the boundary as the loci of pixels which had the
mean intensity in the window with the largest gradient.

The nebulosity might extend to a larger area than the boundaries we
have defined.  However, the intensity in these regions is very low
compared to that measured inside the area chosen for the
integration. We estimate that no more than 10\% of flux is
excluded. We computed the total flux by setting a lower threshold
based on the gradient and a higher threshold based on saturation
limits, and then counting the total number of pixels that were
spatially within the borders. Because of the flux in the very low S/N
area as well as other unavoidable uncertainties, the measurements in
Table 1 can be considered lower limits, while the total flux
might be up to 20\% higher.

Fig.~\ref{fig2} shows the comparison between the bright knots in the
South-West boundary of the nebula images at different epochs. We give
integrated fluxes at the different epochs in Table 1.

\begin{figure}
\resizebox{\hsize}{!}{\includegraphics{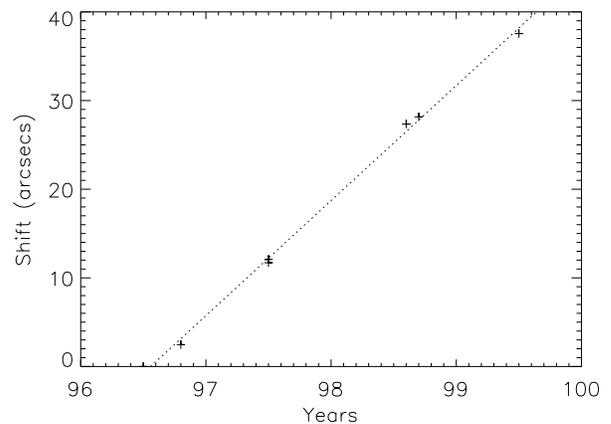}}
\caption{The displacement of the southwestern ``boundary''
of the nebula between May 1996 and July 1999.}
\label{fig3}
\end{figure}

We obtained astrometric information on the clearly visible knot south
of the nebula and on the boundary itself, defined as before. The
position of the area in the knot with maximum counts (averaged on a
10x10 pixel box) was measured. The boundary was fitted with a straight
line. The reference frame was set using three stars (whose centers
were determined with a gaussian fitting). Fig.~\ref{fig2} shows
positions of the nebular front in June 1996 (upper left), June 1997
(upper right), August 1998 (lower left) and July 1999 (lower
right). The positions of the nebular boundary and of the knot give the
same speed. Fig.~\ref{fig3} is a plot of the displacement against time. It
shows that the motion was almost constant (0.034\as{}/day) over three
years, which projects to 0.35 $c$ at an assumed distance of 1.8
kpc. This contrasts with the rate of expansion ($\approx$c) found by
Rosino et al. (1996) that most likely continued between December 1995
and our first observations in May 1996.

\section {Spectroscopy of the nebula\label{spec}}

In November 1998 we took three sets of 90 spectra 
of the bright knot shown in Fig.~\ref{fig2} and 25 loci northwards
and southwards of it. We used the Hydra multi-fiber positioner with a
bench spectrograph and the DensePak fiber optic bundles at the WIYN
telescope. The bundles scanned an area extending about 35\as. Because
of difficulties in the absolute calibration of each individual fiber,
we cannot be precise about the relative intensities of the lines, but
there is little variation in them. The various spectra differ in flux.
The line widths in the nebula are FWHM $\sim$5.9
\AA, perhaps slightly larger than those of the night sky (FWHM
$\sim$5.4 \AA), implying an upper limit of thermal FWHM of 100 km
s$^{-1}$.

\begin{figure}
\resizebox{\hsize}{!}{\includegraphics{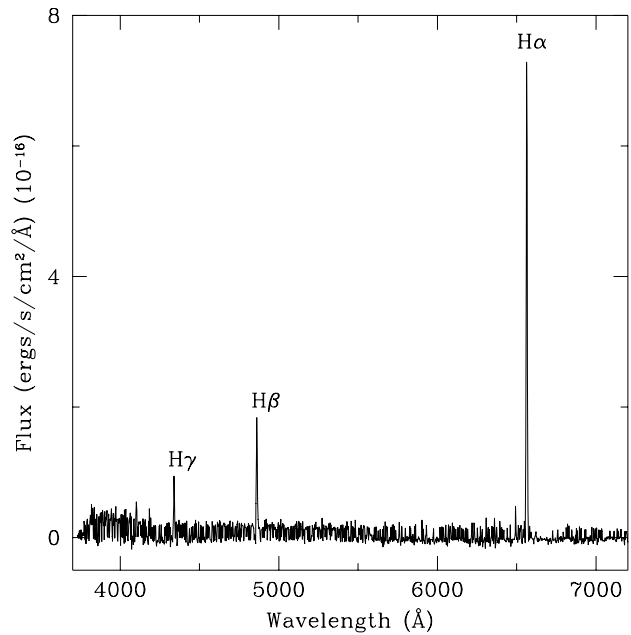}}
\caption{The spectrum taken with the RC
spectrograph on the Kitt Peak National Observatory 4m telescope
on October 2, 1999.}
\label{spectra}
\end{figure}

A year later, on October 2, 1999, a spectrum of the region around Nova
Cyg 1992 was taken with the RC spectrograph on the Kitt Peak National
Observatory 4m telescope.  The KPC-10A grating was used with a 4000
\AA \ width, 2\as{} x 300\as{} slit with 7 \AA \ resolution. The
useful spectral range is from about 3700 to 7500 \AA.  Two 750 s
spectra were taken. Their sum is shown in Fig.~\ref{spectra}.

Because spectral features of the nova are located across the entire
slit, we used a scaled spectrum of another object to remove sky
features. We then extracted the spectrum, wavelength calibrated it
with an FeAr comparison spectrum, and flux calibrated with the star
BD+28 4211. Since we are unable to do an exact sky subtraction, the
absolute calibration is only approximately correct, but the relative
fluxes between the hydrogen emission lines are secure.  The line flux
ratios derived are 4.3 for H$\alpha$/H$\beta$ and 2.1 for
H$\beta$/H$\gamma$. If these lines are produced by recombinations, as
we believe them to be, these ratios imply an interstellar extinction
of $A(V)$ = 0.8, which is reasonable.

As already noted by Garnavich \& Raymond (1997), the nebular spectra
{\em shows only the Balmer lines series, with no other lines present,}
in both our observations taken with the WIYN Densepak fiber array in
November 1998 and in the long slit 4-m spectrum from Sept 1999. The
ratio [N~II]$\lambda$(6583)/\ha\ $<$0.03 (3 $\sigma$). For
comparison, its value is 0.14 in the Orion Nebula (Esteban et
al. 1998), 0.18 in M~17(Esteban et al. 1999), and $\ge$0.4 in the
brightest diffuse galactic emission (Haffner et al. 1999). The upper
limit to He~I $\lambda$5876/H$\alpha$ is 0.03 (3
$\sigma$).
The spectrum is completely different from that of the nova
itself at any stage in its history. The nova had a strong continuum in
its early stages and strong, broad permitted lines of H~I, N~II,
N~III, and Fe~I in its early stages (Rafanelli et al. 1995) and
forbidden lines of [O~III], [Ne~III], [Ne~V], and He~II later (Rosino
et al. 1996). Therefore, the {\it reflection} nebula hypothesis
can be ruled out.

Some parts of the nebula definitely faded with time. An obvious
example is the patch just to the right and below the reference star in
Fig.~\ref{fig2}. It was bright in the first frame (July 1996) and had
disappeared by the last (July 1999).
The production of a spectrum that lacks forbidden lines, and especially
collisionally excited emission from low ionization potential species
such as S$^{+}$ (IP$=$10.36~eV), requires most unusual conditions in the
nebula.

\section{Geometry of the nebula}

If there were a uniform spatial distribution of interstellar material
surrounding the nova, we would see the portion of the nebula that is
ionized by the flash to be expanding at speed $c$. 
However, we have shown that in 1996 -- 1998
bright regions appeared as the wave swept past them with an apparent
rate of expansion of 0.35$c$. This indicates that dense material of
the ISM, which appears brighter than the surrounding tenuous gas after
illumination, is not uniformly distributed in space. There is a sheet
of dense material inclined to the line of sight such that the
projected speed of the illuminated region is reduced. There is so
little absorbing material between the sheet and the nova that the
pulse of radiation reaches it with considerable intensity.

We can determine the distance from the star of newly illuminated
material by using its angular displacement and the time delay of the
arrival of its radiation. The apparent rate of expansion of the
boundary gives the difference in distances along the boundary, or
equivalently the angle of the boundary relative to the line of sight
or to the nova.

The geometry is shown in Figure 5. The observer is assumed to
be at distance $D$ vertically upwards in the figure. Consider the
point $P$ at the edge of the nebula, which begins to emit its
radiation when the pulse of ionizing radiation from the nova
arrives. The intensity along the line of sight to $P$ increases
abruptly when the ionizing flash reaches $P$. Let $\Delta t$ be the
time interval between the arrivals of the light to us from the nova
and from $P$ (this is not the same as the time since the explosion,
since $P$ is not at the same distance as the nova). The distance of
$P$ from the nova is $r$ and its projection along the line of sight is
$s$. Then we have
\begin{eqnarray}
 c\,\Delta t =&r-s=r-(r^2-\theta^2D^2)^{1/2}\:; \nonumber\\
r=&(c\Delta t/2)\,\left[1+\theta^2D^2/
(c^2\Delta t)^2)\right]\:, \label{reqn}
\end{eqnarray}
\noindent where $\tan\,\alpha=s/(\theta D)$.

Consider the bright knot with \th\ = 3.1\am{} in July 1998,
6.4 years after the explosion. Then \cdt\ = 6.4 light years (ly) and
\thd\ = 5.3 ly, yielding $r$ = $5.4\pm0.4$ ly, $s=1.0\pm0.5$ ly, and
$\alpha \approx 10^o $. Thus, the knot is almost the same distance as the
nova. The errors in $r$ and $s$ are based on the uncertainties in
\cdt\ and \thd. The uncertainty of \cdt\ arises from guessing when the
ionizing pulse broke out of the nova's shell, relative to the
beginning of the brightening of the star. Another contribution is the
length of the ionizing pulse, since the knot integrates the incident
ionizing radiation over time. The uncertainty of \thd\ arises from the
error in $D$, which is $\approx0.1D$. Our guess at the spatial
thickness of the wave of ionizing photons is 3 light months, and that
the flash was emitted relatively soon after the outburst. The
justifications of these assumptions will be given in Section 7.2.

\begin{figure}
\resizebox{\hsize}{!}{\includegraphics{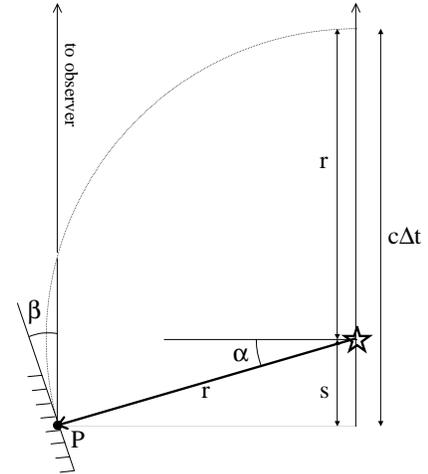}}
\caption{The assumed geometry of the material in the ``bright
knot'' shown moving in Fig.~\ref{fig2}  ionized 
by the nova (the star in the diagram). The observer is
at distance D in the upward direction. Light from the point $P$, seen
at angular separation $\theta$ from the nova, reaches us at time
$\Delta t$ after the direct light from the nova. Point $P$ is on the
front face of the  dense region (defined
as the ``knot'' and the material inside the
``boundary'' in the text) that shows an apparent expansion of the
advancing wave at a rate $d(\theta D/d(\Delta t)$. The face is
inclined at the angle $\beta$ to our line of sight.}
\label{geom}
\end{figure}

The rate of expansion of the bright edge of the nebula provides an
estimate of the tilt of the edge with respect to the line of sight. We
have a diagram very similar to Fig.~\ref{geom}, but considering two
points and the difference in time of arrival of their radiation. Let
$\beta$ be the angle of their separation (i.e., of the leading edge of
the knot) relative to the line of sight, measured away from the
direction of the nova. Then some geometry yields $\beta$ in terms of
the rate of expansion of the nebula:
\begin{equation}
d(\theta D)/d(c\Delta t)=\sin(\beta)/\left[\sin(\beta-\alpha)
+\cos(\beta)\right]\:.\\
\end{equation}
\noindent With $ d(\theta D)/d(\Delta t)$ = 0.35$c$, the expansion
speed of the nebula, $\beta=(25\pm7)^o$.

\section{Search for other nebulae}

We searched for H$\alpha$ emission in the fields of other recent and
old novae. We chose a sample that were bright in outburst and are
still relatively bright at quiescence, and which exploded at very
different epochs. We included two recurrent novae. We observed Novae
V603 Aql 1918, DK Lac 1950, HR Del 1967, QU Vul 1984, Sgr 1993, Cas
1993 and 1995 and the recurrent nova RS Oph with the R and H$\alpha$
filters with the WIYN telescope. The field of view was 6.8\am{}.  In
all cases we set upper limits on the order few times 10$^{-18}$ erg
cm$^{-2}$ s$^{-1}$ arcsec$^{-2}$ for the H$\alpha$ intensity.  With
the 1m REOSC reflector at the Torino Observatory we observed 11\am{}
radius fields around Novae V1500 Cyg 1975, CT Ser 1948, U Leo 1855,
and the recurrent nova T Cr Bor. In this case, the upper limits on the
H$\alpha$ flux were two orders of magnitude higher. While the results
of this survey will be presented more in detail in a forthcoming
paper, we want to stress that the nebula around N Cyg 1992 must be a
rare phenomenon and the product of special conditions.

\section{Explaining the spectrum}

Our spectra of the nebula show only Balmer recombination lines. It is
almost impossible to believe that the bright material is poor in heavy
elements relative to the ISM in general. If anything, previous nova
outbursts from the system should have slightly enriched the
circumstellar medium.  The lack of the collisionally excited lines in
the spectrum can be understood in two ways:

(a) The nebula has been photoionized by Lyman continuum radiation from
the nova, similar to ordinary H~II regions except that the excitation
is much lower and the nebula is not in a steady state. In this case,
\te, the electron temperature in the ionized nebula, must be
relatively low ($\le$3300 K), thereby suppressing the collisional
forbidden lines and increasing the emissivity of the Balmer lines.

(b) Perhaps the nova released a flood of Lyman line emission that
impinged upon the neutral nebular cloud. In this case, the Balmer
lines are produced by fluorescence following absorption and subsequent
cascades of the Lyman line photons.

We consider the Lyman-line fluorescence scenario very unlikely. The
observed Balmer decrement is remarkably like recombination with a
modest and plausible amount of interstellar reddening ($A(V)\:\sim$
0.8 mag). By contrast, it seems difficult to produce this spectrum
with Lyman lines incident from the outside of a slab. The Lyman
decrement would heavily favor Ly$\beta$ over Ly$\gamma$ and higher
members of the series because of the scattering of the lines in both
the nova shell emitting them and in the cold nebula absorbing
them. Following each scattering there are conversions into Ly$\beta$
and a Paschen photon. This process is why ``Case B", in which the
Lyman lines are all converted into Ly$\alpha$ plus members of higher
series, is adopted in almost all standard nebular
analyses.\footnote{In the case we consider, Ly$\alpha$ is not relevant
because it cannot produce a Balmer line following its absorption.}
Another difficulty with this scenario would be that it is difficult to
suppress the Lyman continuum radiation from the nova shell if the
Lyman lines escape readily. The escaping Lyman continuum would ionize
and heat the nebula in exactly the way we suggest in the recombination
scenario outlined above. Still another problem is that the absorption
of the Lyman lines would be very inefficient because the width of the
nova lines would be $\sim$2100 km s$^{-1}$ (Rafanelli et al. 1995) in
the early stages and 1600 km s$^{-1}$ somewhat later.
By contrast, the absorption lines
in the neutral material are observed to be relatively narrow (FWHM
$\le$ 100 km s$^{-1}$. They probably have a thermal plus turbulent
width of 10 -- 20 km s$^{-1}$, typical for the ISM. Thus, the
absorption of Lyman lines by the neutral material would be very
inefficient. Each Lyman line requires almost as much energy as a Lyman
continuum photon, so this inefficiency greatly increases the energy
required from the nova. Lyman continuum photons produce narrow
recombination lines regardless of the velocity spread of the continuum
source.

\subsection{Physics of flash ionization}

The very strange spectrum of the nebula -- Balmer emission lines,
together with the absence of forbidden lines -- requires very unusual
conditions. The forbidden lines cannot be wholly suppressed by
de-exciting collisions without encountering physical
inconsistencies. The brightest forbidden line expected is
[N~II]$\lambda$6563. Its ``critical density", at which the rates of
collisional and radiative de-excitations from the upper level are
equal, is $n_{\rm cr}\, =\, 6.3\times10^4 \,(T/5000\, {\rm K})^{0.5}$
cm$^{-3}$. Suppressing it below its observed upper limit requires 10
times this density if $T$ = 5000 K and 200 times at $T$ = 10000 K. At
these densities, recombination of H would take place within 40 days
for $T$ = 5000 K. In order to suppress the forbidden lines, we require
two conditions: (a) a rather soft ionizing spectrum, and (b) a
substantial fraction of neutral H atoms mixed with the ionized.

When the ionizing radiation impinges upon the cold neutral nebula, it
creates photoelectrons that collide with their neighboring electrons,
protons, and neutral atoms, with timescales decreasing in that order.
The collisional processes are discussed in Spitzer (1978). All such
times are short: minutes to a few days; the slowest is the exchange
between H$^+$ and H$^0$. Thus, there will be only one temperature
shared by all particles.

The incident ionizing radiation must be relatively soft so that the
ejected photoelectrons do not have excessively large energies after
ionizing the H$^0$. We envision the incident radiation to be
distributed between the ionization edges of H (13.6 eV) and of He$^0$
  (24.6 eV), or possibly of Ne$^0$ (= 21.6 eV). The softness of the
incident spectrum is shown by the observed absence of any He~I
recombination lines in the spectrum, indicating that He is much less
ionized than H. Presumably, photons at higher energies were absorbed
by the large amounts of Ne, He, and O that were present in the ejecta,
preventing the escape of their ionizing radiation.

\subsection{A partially ionized nebula}

In a highly ionized nebula, the photoionization heating from the
ionizations will be shared between only the ions and electrons. If the
ionizing spectrum extends up to even 22 eV, a fairly low threshold,
the nebula will be hot and the usual optical forbidden lines will be
strong. For instance, if the mean energy of an ionizing photon is 18
eV (as for a uniform photon intensity between 13.6 and 22 eV), the
mean kinetic energy of each particle, including He$^0$ but neglecting
H$^0$, is 2.0 eV, and $T\,\sim\,23000$ K. The [N~II] and [S~II] would
be comparable to \ha. Lowering [N~II]$\lambda$6583/\ha\ to 0.030, 3
$\sigma$ above the observed limit, requires $T\le$ 3300 K.

We must appeal to {\it partial} ionization of the nebula. The low \te\
can be achieved by having H$^0$ atoms share the deposited
energy. He$^0$ is far from sufficient to provide these neutrals. The
required fraction of H$^+$, $q$, depends rather sensitively on
$\langle h\nu\rangle$, the mean energy of the ionizing photons,
because the mean photoelectron energy is ($\langle h\nu\rangle -13.6$
eV). If $\langle h\nu\rangle$ = 18 eV, $q\, \sim \,0.08$; if $\langle
h\nu\rangle$ = 16 eV, $q\,\sim \,0.13$ because fewer neutrals are
needed to share the energy.

The low level of ionization can be achieved when a burst of ionizing
radiation impinges on dense neutral material. Since the
photoionization cross section $\propto\nu^{-3}$, the harder radiation
penetrates more deeply into the neutral gas and deposits the more
energetic photoelectrons into regions with more neutrals to share
their energy. With sufficient neutral density, the ionization fraction
can be made low enough to avoid high temperatures.

The electron (or H$^+$) density is poorly known. The mean \ha\
intensity of $1.0\times10^{-3}$ erg cm$^{-2}$ s$^{-1}$ sr$^{-1}$
within the bright knot of emission 30$\arcsec$ across shown near the
bottom of the lowest frames in Fig.~\ref{fig2}. This translates to an
emission measure of $4.0\times10^3\,(T/3300 {\rm K})^{0.91}$ cm$^{-6}$
pc, or $n_e\sim$ 120 cm$^{-3}$ in the ionized part of the sheet.

We feel that $n_e\sim$ 120 cm$^{-3}$ is a severe lower limit. Let us
calculate how many surfaces of ionized material would be required to
provide the observed average intensity of \ha\ within the bright
patch. The thickness, $L$, of the ionized region normal to the radius
to the nova (not along the line of sight) is $L=[n({\rm H}^0)
\langle\sigma({\rm H})\rangle]^{-1}$, where $\langle\sigma({\rm
H})\rangle$ is the absorption cross section of H averaged over the
ionizing spectrum. For $\langle h\nu\rangle = $18 eV, this gives L =
3.4$\times 10^{17}$/n(${\rm H}^0$) $\theta$, cm, subtending only
0.12$\arcsec$ at 1800 pc if $n^0$ = 100 cm$^{-3}$. Our spectra and
images might contain the emission from many unresolved surfaces, as is
suggested by the rippled structure in the nebula seen in
Fig.~\ref{fig1} and ~\ref{fig2}. The \ha\ intensity normal to each
nebular surface is $I({\rm H}\alpha)=n_en_p\,j({\rm H}\alpha)L$, where
$j({\rm H}\alpha)$ is the emissivity of \ha.  Each surface produces an
emission measure of $0.11\,n\,q^2/(1-q)$ cm$^{-6}$ pc in the direction
normal to it, where $n$ is the total H density. We have estimated
above that $q\sim0.1$ in order to provide the low temperature. If
$n\sim10^3$ cm$^{-3}$, so that $n_e\sim$ 100 cm$^{-3}$, we would
require $\sim 4000$ sheets of emitting material to provide the
observed \ha\ along the line of sight! If we assume that no more than
a few surfaces are pierced, say 20 ripples each pierced twice, the
$n_e$ becomes $\sim10^4$ cm$^{-3}$. This large density, needed to
explain the highest surface brightness, has only a very marginal
effect on [N~II]$\lambda6563$.

Interstellar extinction is not a barrier to seeing through
several folded sheets of material. The absorption cross section of
H$^0$ at the Lyman edge is $1.3\times10^4$ times the extinction cross
section at $V$. The nebular sheets can have very a low optical depth
to dust at all wavelengths.

Let us consider cooling following the deposition of the kinetic energy
of the photoelectrons into the gas. Dynamical effects have a
timescale, $t_{\rm dyn}$, of $L/c_s$ ($c_s$ is the adiabatic speed of
sound) = $2\times10^4/n({\rm H}^0)$ years at 3300 K. At the very large
densities we estimated above, $t_{\rm dyn}\sim1$ year. In this case,
part of the cooling will arise from the expansion of the newly ionized
matter and part from radiation. The main contributions to the
radiative cooling are from free-free emission and
recombinations. Collisional cooling from far-infrared fine structure
lines is largely suppressed by the high $n_e$, and the optical
forbidden lines by the low $T$ (by design). The main forbidden lines
will be [N~II]$\lambda$6583 and [S~II]$\lambda$6717, the dominant
species of their respective elements independently of the level of
ionization of H.

The cooling processes in the nebula are not able to suppress the
forbidden lines within the timescale of a year.	 Without the charged
particles sharing the liberated photoelectric energy with neutrals,
the expansion would have to cool from about 15000 K to 3300 K, a
factor of 4.5, requiring an expansion by a factor of 4.5$^{1.5}$ = 10,
requiring a few times $t_{\rm dyn}$. Most of the emission would arise
in the dense regions that have not had time to expand ($j\propto
n^2$). Dynamical effects cannot suppress the optical forbidden lines
until more than a year after the flash unless there is a large fraction
of neutrals in the plasma. The radiative rate of cooling for $\langle
h\nu\rangle$ = 18 eV, neglecting expansion, is $\sim100$ K yr$^{-1}$,
so the total radiative cooling of the nebula since the flash is not
large.

The very high density needed to avoid too many emitting surfaces leads
to a recombination timescale of $(n_e\langle\sigma({\rm
H})\rangle)^{-1}\sim4$ years, comparable to the dynamic timescale,
since there is not a continuous supply of ionizing photons.

The timescale of the ionizing flash was probably two or three months.
The nova atmosphere started cold ($\sim$ 20000 K) but reached very
high temperatures ($> 5 \times 10^5$ K) a few months later (Balman et
al. 1998). Since the initial bolometric luminosity was almost constant
at $1.6\times10^{38}$ erg s$^{-1}$, the radius of the star was
decreasing rapidly as $T$ increased ($R\propto T^{-2}$). The result of
the rapidly decreasing radius and shift of the spectrum to higher
energies is a rapid drop in the luminosity in H-ionizing photons. The
ionizing burst was, presumably, emitted while its temperature was
$\sim$ 35000 -- 60000 K, when the peak of the
blackbody emission is in the range of H ionization energies. The
observed very hard photons (in the supersoft X-rays phase at maximum)
must have also produced some ionization in the surrounding medium, but
the relative numbers of the softer H-ionizing photons was probably
vastly greater.

The total energy required in the ionizing burst is\\
 $4\pi d^2N\langle h\nu\rangle$, where $N$ is the column density of ions or electrons in
a slab perpendicular to the radius from the nova and $d$ is the
distance of the slab to the nova, 5.4 ly for the patch we discussed in
\S5. The column density is the local density, $n_e$, times the path
length, $[n_0\,\langle\sigma({\rm H})\rangle]^{-1}$. This gives
$N\,=\,q/[(1-q)\langle\sigma({\rm H})\rangle]$.	 With $\langle
h\nu\rangle$ = 18 eV, $N\sim4\times10^{16}$ cm$^{-2}$.	The
total energy of the burst, if it was isotropic, was
$\sim4\times10^{44}$ erg, or the bolometric
luminosity of the nova for $\sim$4 weeks. This is, obviously, a very
substantial energy requirement. It could be avoided by assuming
clumping in the nova envelope so that the ionizing radiation covered
less than $4\pi$ steradians. The explanation of the Balmer lines by
fluorescence of Lyman radiation would require much more energy because
of the inefficiency of absorption of the wide Lyman lines.

The mass of the illuminated portion of the nebula is $M_{\rm neb} =
4\pi d^2 \mu m_{\rm H}/\langle\sigma({\rm H})\rangle$, where $\mu$ (=
1.4) is the mean mass per H nucleus. Then $M_{\rm neb}\,=\,0.13
M_\odot$ for $\langle h\nu\rangle$ = 18 eV and 0.10 $M_\odot$ for 16
eV. This could be material compressed by a previous outburst. The mass
included in a 5 ly sphere is $\sim 0.4 n_H$(ISM) \msol, so we would
expect to find the required nebular mass from the ISM in the Cygnus
spiral arm. Its composition should be heavily dominated by the ISM
material swept up by a previous expanding shell.

If the dense sheets are the results of the ``snowplow" phase of a
previous outburst, the radial momentum in the previous outburst equals
the nebular mass times the turbulent speed (reflecting the pressure)
of the shell;\\
 $M_{\rm nova}\,v({\rm shell}) = M_{\rm nebula}\,v({\rm
turbulent})$. The present outburst consisted of (2 --
5)$\times10^{-4}$ \msol\ (Woodward et al. 1997) at 2000 km s$^{-1}$.
If we assume that it was typical, the snowplow phase would end when $V
({\rm turbulent})$ = (4 -- 10) km s$^{-1}$, which is
satisfactory. However, the illuminated nebular mass is near the upper
end of the value that can be explained, and we would expect that much
of most of the nebular material has not been illuminated.

Compressing of the ISM into the dense sheets by the snowplow phase
does not explain why this object is so unusual, although V1974 Cyg has
exceptionally massive ejecta. Another possibility is that the sheets
are the results of the ejection of a common envelope binary, in which
mass exchange occurs so fast that the acceptor star cannot accrete the
flow either directly or in a disk. A supergiant atmosphere forms with
the binary stars contained deep within its interior. The atmosphere
(up to 1 \msol\ or more) is ejected at speeds near the escape speed
from a supergiant (10 km s$^{-1}$). This material has much more
outward momentum than the nova ejecta.

The common envelope scenario would surely be a very unusual
event. Another advantage of this explanation is that the nebular mass
would be expected to be much larger than just the visible, since it
would include the ISM that was swept up in addition to that in the
common envelope.

\section{Summary }

The region surrounding Nova Cyg 1992 shows an expanding region of
brightened material that has the following observed characteristics:

1. The speed of expansion of the outer edge of the illuminated region
was initially close to $c$, but slowed to 0.35$c$ by 1996. This
clearly shows that the illumination is a wave of radiation expanding
from the nova reaching surrounding material.

2. Spectra of the nebula taken over a period of 3 years with moderate
resolution show only Balmer lines that have a recombination spectrum
with moderate reddening. The 3$\sigma$ upper limits to [N~II], [S~II],
and He I $\lambda$5876 relative to \ha\ are $\sim$0.03. A spectrum with
the KPNO 4-m telescope showed that the lines are narrow ($\le 100$ km
s$^{-1})$ compared to the nova expansion speed (2000 km s$^{-1})$.

3. Regions of the bright nebula have faded over a timescale of 2
years.

Our interpretation of the observations is based on an adopted distance
of ($1800\pm200$) pc. At this distance, the bright patch for which we
have the best spectrum is 5.4 ly from the nova and $\sim$1.0 ly more
distant from it.

We interpret the spectrum as recombinations of ionized H in a plasma
of normal ISM composition. An alternative possibility is fluorescence
following absorption of a burst of Lyman line radiation from the nova.
We consider this very unlikely because (a) the spectrum would have a
steeper \ha/H$\beta$ ratio than observed, (b) it would be difficult for the
nova shell to emit higher Lyman lines copiously without accompanying
Lyman continuum, and (c) the fluorescence would be very inefficient
because of the great velocity width of the Lyman lines from the nova.
The energy requirement for the burst would much larger than if the
emission were in the Lyman continuum.

It is most unlikely that the illuminated material is strongly lacking
the heavy elements that produce forbidden optical lines. The density
will lead to collisional de-excitation of the various infrared fine
structure lines and [S~II]$\lambda6717$, but not
[N~II]$\lambda6583$. A very low temperature, $\sim3300$ K, is required.
The low $T$ is achieved by a sharing of the kinetic energy of the
ejected photoelectrons with neutral H atoms that dominate the ion
density of the partially ionized plasma by about a factor of
ten. These neutrals serve as a reservoir of energy that delays cooling
beyond the recombination time.

The electron density is poorly determined. We can predict the
intensity of \ha\ in the direction normal to each sheet and then
account for the maximum observed surface brightness. The electron
density must be very high ($\sim 10^4$ cm$^{-3}$ unless our line of
sight pierces thousands of emitting surfaces. The extinction by dust
in the sheets is not important.

The ionizing radiation was probably emitted when the temperature of
the nova was in the range 35000 -- 60000 K, when the maximum emission
is in the range of H ionization. The nova was observed to start rather
cool and reach 200000 K after 6 months. The energy required by the
ionization radiation, if the emission was isotropic, was the
luminosity of the nova times about four weeks, barely within the
plausible range. The energy requirement is much greater if the Balmer
lines are produced by fluorescence of Lyman lines.

At the very high densities we envision, the dynamical timescale is about
a year, and the recombination timescale is comparable. The fading of
parts of the nebula within 2 years is caused by recombinations
and expansion after the burst had passed.

The mass of the ionized sheets is about 0.1 \msol. This can barely be
supplied by the snowplow phase of the previous nova ejection if it was
as massive as the present one. Alternatively, it could easily be
supplied if the previous phase was a common envelope binary, in which
the mass exchange is so rapid that the stellar orbit occurs within a
common atmosphere that is expelled at about the escape speed. This
very rare event would supply more than enough momentum to compress the
expelled material and surrounding ISM into sheets of neutral material.

We have also undertaken sensitive surveys of H$\alpha$ nebulae around
recent novae but find no evidence of other such nebulae, so this type
of object must be rare.

\begin{acknowledgements}
This research was initiated by the late Prof. Leonida Rosino, who was
a teacher and a mentor for two of us (M.O. and M.D.V.).  The work was
supported by the Italian Space Agency ASI and by the Italian MURST
40\% program (``cofinanziamento osservatori-universit\`a''). We thank Annette
Ferguson for allowing us to take the spectra with the Kitt Peak 4m
telescope during her observing run, and Samar Safi-Harb for her help
in the early stage of the imaging program. We are also grateful to
Takashi Ijima for discussing the early spectra with us and to Marcella
Contini for  critical reading of an early draft. 

\end{acknowledgements}

\end{document}